\documentclass{article}
\usepackage{amsmath}
\usepackage[numbers,sort&compress]{natbib}
\usepackage{notoccite}
\usepackage[utf8]{inputenc}
\usepackage{gensymb}
\usepackage{graphicx}
\usepackage{dblfloatfix}
\usepackage{authblk}

\usepackage{geometry}
\usepackage{color}

\title{Planar graphene-NbSe$_2$ Josephson junctions in a parallel magnetic field} 

\author[1]{\thanks{Equal contribution}Tom Dvir}
\author[1]{$^{\ast}$Ayelet Zalic} 
\author[2]{Eirik Holm Fyhn}
\author[2]{Morten Amundsen}
\author[3]{Takashi Taniguchi}
\author[4]{Kenji Watanabe}
\author[2]{Jacob Linder}
\author[1]{Hadar Steinberg}
\affil[1]{\textit{The Racah Institute of Physics, The Hebrew University of Jerusalem, Jerusalem 91904, Israel}}
\affil[2]{\textit{Center for Quantum Spintronics, Department of Physics, Norwegian University of Science and Technology, NO-7491 Trondheim, Norway}}
\affil[3]{\textit{International Center for Materials Nanoarchitectonics, National Institute for Materials Science, 1-1 Namiki, Tsukuba 305-0044, Japan}}
\affil[4]{\textit{Research Center for Functional Materials, National Institute for Materials Science, 1-1 Namiki, Tsukuba 305-0044, Japan}}

\date{}

\begin{document}

\maketitle

\begin{abstract}
\textbf{Thin transition metal dichalcogenides (TMDs) sustain superconductivity at large in-plane magnetic fields due to Ising spin-orbit protection which locks their spins in an out-of-plane orientation.
Here we use thin NbSe$_2$ as superconducting electrodes laterally coupled to graphene - making a planar, all van der Waals (vdW) two-dimensional Josephson junction (2DJJ). We map out the behavior of these novel devices with respect to temperature, gate voltage, and both out-of-plane and in-plane magnetic fields.
Notably, the 2DJJs sustain supercurrent up to parallel fields as high as 8.5 T, where the Zeeman energy $E_Z$ rivals the Thouless energy $E_{Th}$, a regime hitherto inaccessible in graphene.
As the  parallel magnetic field $H_\parallel$ increases, the 2DJJ's critical current is suppressed, and in a few cases undergoes a suppression and recovery. 
We explore the behavior in $H_\parallel$ by considering theoretically two effects: A 0-$\pi$ transition induced by tuning of the Zeeman energy, and the unique effect of ripples in an atomically thin layer which create a small spatially varying perpendicular component of the field. 
2DJJs have potential utility as flexible probes for two-dimensional superconductivity in a variety of materials, and introduce high $H_\parallel$ as a newly accessible experimental knob. }
\end{abstract} 

By coupling graphene to exfoliated superconductors such as NbSe$_2$~\cite{Efetov_2016, Sahu_2018, Lee2019} it is possible to realize Josephson junctions where both the normal and superconductor materials are two-dimensional (2D).  
Such junctions should sustain high in-plane magnetic fields.
Thin NbSe$_2$ retains superconductivity at very high in-plane fields due to a combination of suppressed orbital depairing and Ising protection against pair-breaking~\cite{Xi_2016,Dvir_Nat_Comm_2018}, and can sustain magnetic fields above 8 T without any measureable effect on the gap size~\cite{Xi_2016,DvirPRL}.
Coupling graphene to two NbSe$_2$ flakes results in an all van- der-Waals (vdW) two dimensional Josephson junction (2DJJ).
The response of such 2DJJs to in-plane magnetic field will be dictated by both spin and orbital effects. 
In the graphene layer, forming the weak link, the response of carrier spins to the Zeeman field may lead to interesting phenomena such as finite momentum Cooper pairing and a 0-$\pi$ transition~\cite{Linder2008, Liang_2008_0_pi, Moghaddam_2008_F_Gr,Asano_2008}. 
However, the deviation of such devices from the ideal 2D geometry due to ripples and other deformations is significant, as it gives rise to field components perpendicular to the local sample plane, introducing orbital dephasing. 
The latter also occurs due to the bending of magnetic field flux lines, which cannot be considered truly parallel, as they are deflected by superconducting leads~\cite{Suominen2017}. 

2D Josephson devices are a useful platform for the study of finite momentum superconducting states:
Cooper pairs may survive in the spin-polarized Fermi surface created at high magnetic fields by attaining a finite center-of-mass momentum~\cite{Fulde_Ferrel_1964,Larkin_Ovchinnikov_1965}, which translates into a spatially varying order parameter. Finite Cooper-pair momentum $q = 2E_Z/\hbar v_F$ is dictated by the Zeeman energy $E_Z=0.5g \mu_B H$, where $g$ is the Land\'e factor, $v_F$ is the Fermi velocity, and $\mu_B$ is the Bohr magneton. The resulting oscillation of the order parameter within the junction can create $\pi$-phase junctions, where the transition to the $\pi$-phase is found in junction lengths $L$ determined by the multiples of $\pi/q$. 
Weak links characterized by large $g$-factors have shown signatures of finite momentum Cooper pairing \cite{Chen_2018} and allowed the realization of tunable Zeeman driven 0-$\pi$ transitions~\cite{Hart_2016,Li_2019,Ke_2019}.
Graphene should also exhibit a Zeeman-driven 0-$\pi$ transition~\cite{Linder2008, Liang_2008_0_pi, Moghaddam_2008_F_Gr,Asano_2008}. However, reaching this transition requires the application of high magnetic fields, due to the low $g$ factor which limits the momentum shift of the Cooper pair. 
Ballistic graphene is uniquely expected to produce field-tunable switching between 0 and $\pi$ phases while retaining a finite critical current \cite{Linder2008}, and is expected to exhibit triplet superconductivity~\cite{Linder_Triplet_2010}. 

However, the entirely 2D nature of the graphene sheet gives rise to a unique form of disorder due to graphene ripples in the third dimension. In the presence of applied $H_\parallel$ this introduces a small component of perpendicular field with a disorderly spatial variation created by the ripple pattern. This effect can lead to critical current decay with parallel field, a non-Fraunhofer interference pattern, and suppression and recovery of the critical current mimicking a 0-$\pi$ transition. The effect of ripples changes depending on ripple amplitude and wavelength, and junction dimensions~\cite{Fyhn2020}. 
Thus in any experiment involving graphene in parallel field, or indeed, we believe - any 2D conductor in parallel field - this effect should be considered. 
The morphology and effect of ripples is expected to change depending on the substrate and thickness of the 2D layer. Due to the high parallel fields sustained by the junction, our 2DJJ is sensitive to both long and short wavelength sub-milliradian curvature and sub nm height variation in graphene. 

We study planar NbSe$_2$-Graphene-NbSe$_2$ junctions, fabricated by transferring cracked NbSe$_2$ on exfoliated graphene (see Figure \ref{fig:fab}, and detailed information in Supplementary Section 4 \cite{supp}). The thickness of the NbSe$_2$ flakes used for the devices in this paper was around 5-10 nm, evaluated by optical contrast. The junctions exhibit supercurrent characteristics which are similar to diffusive graphene-based devices fabricated using evaporated superconducting electrodes, including gate-tunable critical current and a Fraunhofer-like interference in out-of-plane field~\cite{Heersche2007,Jeong2011,Komatsu2012,Ke2016,Li2016}.
Upon application of in-plane field, the 2DJJ critical current undergoes exponential suppression and transitions from a Fraunhofer to SQUID-like interference pattern, which is retained as the field is further increased up to 8.5 T. We focus our report on Junction A, with NbSe$_2$ thickness of around 10 nm and a weak link consisting of monolayer graphene.  
In a this device we find that the supercurrent exhibits a pronounced suppression-recovery pattern, a feature which may be associated either with a 0-$\pi$ transition, or with with the effect of graphene ripples.

\begin{figure}[h]
    \centering
    \includegraphics[width=0.9\textwidth]{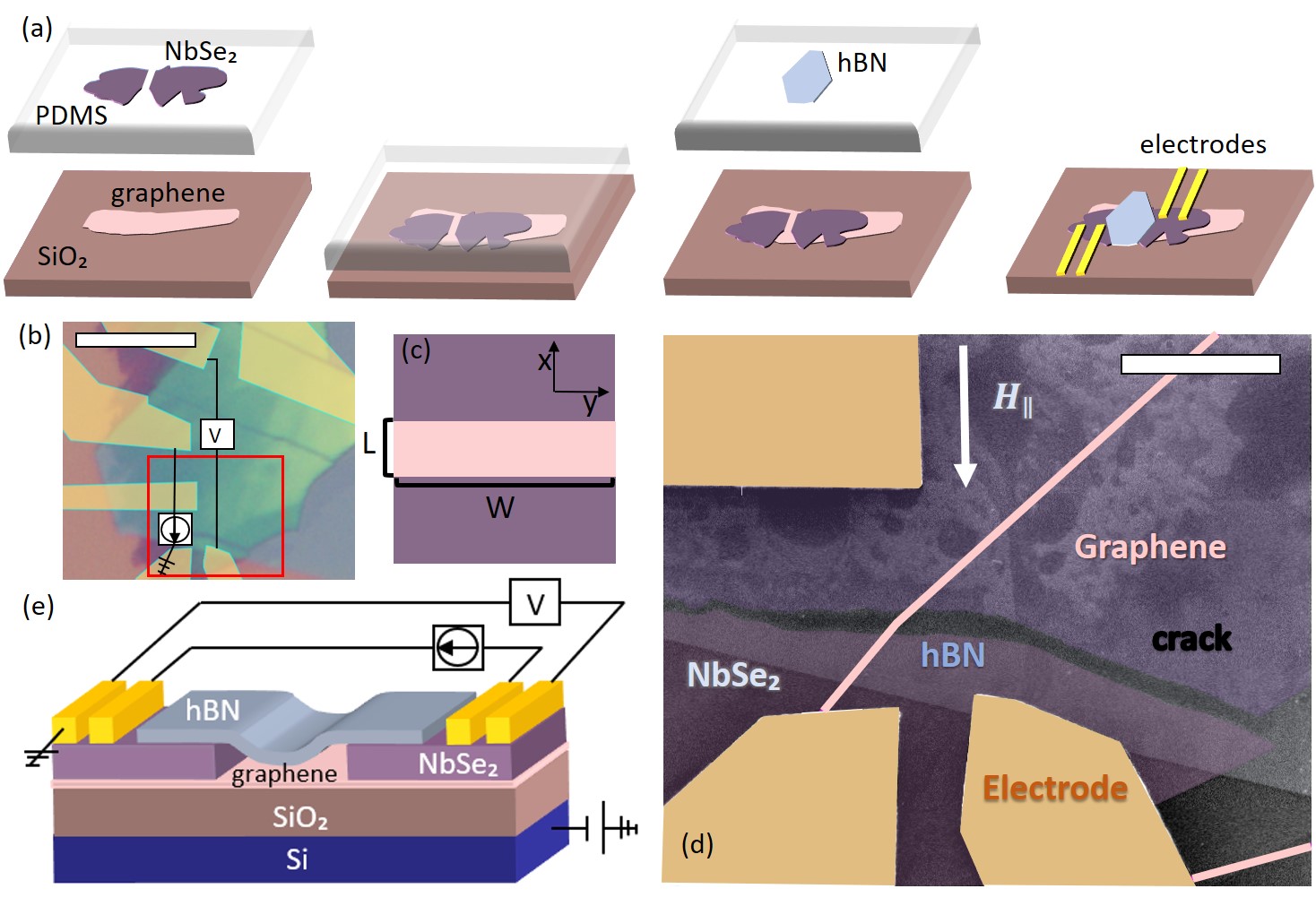}
    \caption{\textbf{a.} Fabrication of planar graphene-NbSe$_2$ JJs involves 1. Exfoliation of graphene on SiO$_2$ and NbSe$_2$ on PDMS. 2. Stamping a cracked 
    NbSe$_2$ flake onto graphene 3. Stamping a thin hBN flake for encapsulation of the crack. 4. Patterning of electrodes. Steps illustrated from left to right \textbf{b.} Optical image of Junction A with schematics of current flow. NbSe$_2$ thickness is around 10 nm. Scale bar 10 $\mu m$ \textbf{c.} Illustration of a rectangular junction geometry \textbf{d.} A false color SEM image of the region marked in a red square from panel b., showing the actual junction geometry, with graphene flake contour highlighted and direction of $H_\parallel$ indicated. Scale bar 2 $\mu m$ \textbf{e.} Schematic illustration of the JJ in a four-probe electronic configuration. Current flows in plane from NbSe$_2$ to graphene to NbSe$_2$. The crack is shielded from the top by hBN. Gate voltage is applied across the SiO$_2$ dielectric.     }
    \label{fig:fab}
\end{figure}

\begin{figure}[!h]
	\centering
	\includegraphics[width=0.9\textwidth]{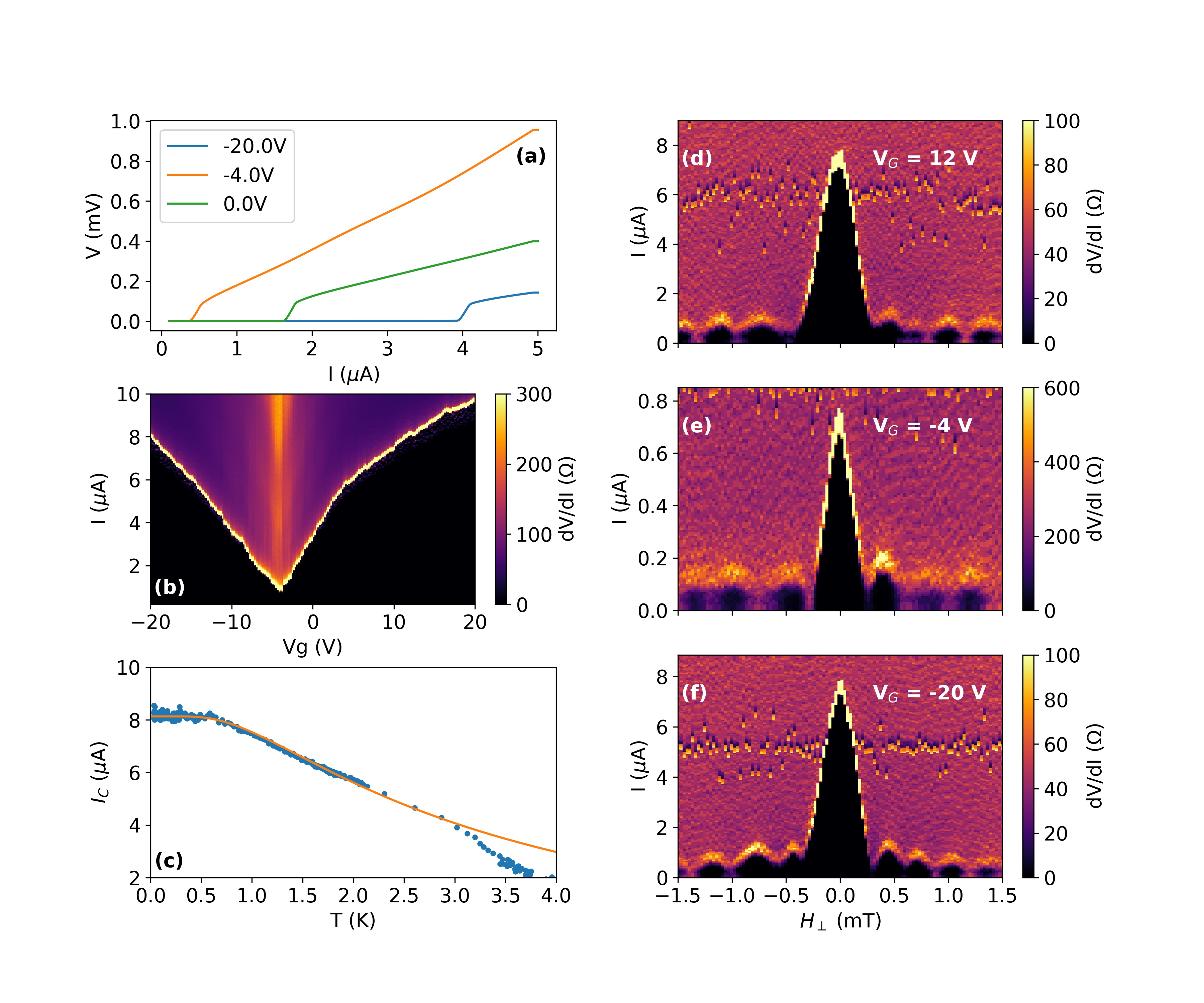}
	\caption{\textbf{a.}  $I-V$ curves of Junction A (monolayer graphene) taken at different gate voltages (see legend). \textbf{b.} Differential resistance ($dV/dI$) of Junction A as a function of bias current and gate voltage. \textbf{c.} Temperature dependence of the critical current of Junction A (blue dots) and a fit to Eq. \ref{eq:temp} (orange line), taken with gate voltage of  -20 V.  \textbf{ d.,e.,f.} Differential resistance of Junction A as a function of bias current and external perpendicular magnetic field, taken with gate voltages of 12 V, -4 V (Dirac point), and -20 V, respectively.  All the panels show data at $H_\parallel = 0$ T and $T = 30$ mK.  }
	\label{fig:JJ}
\end{figure}

We begin by characterizing the transport of a 2DJJ. Figure \ref{fig:JJ}(a) shows the typical current-voltage characteristics of Junction A, where the $I$-$V$ curves at different gate voltages exhibit a switching behavior between zero resistance and finite resistance at the junction switching current $I_C$. Typical to density-tunable graphene JJs \cite{Heersche2007}, $I_C$ is modulated by the gate voltage $V_G$, and reaches a minimal, yet finite, value of $I_C \approx 0.4 \mu A$ at the Dirac point $V_G = -4$ V. This is evident in panel (b), where the differential resistance $dV/dI$ vs. $I$ and $V_G$ is presented as a color plot.
Thus, our 2DJJs exhibit the same bipolar super-current expected in graphene-based Josephson devices \cite{Heersche2007}. 

The Thouless energy, $E_{Th}$, defined as the inverse of the traversal time of the junction, is an energy scale characteristic of normal transport, which also governs the superconducting properties of Josephson junctions \cite{Tahvildar-Zadeh2006}. Josephson junctions vary between regimes defined as long ($\Delta / E_{Th}>>1$) or short ($\Delta / E_{Th}<<1$), and diffusive ($L<l$) or ballistic ($L>l$), where $l$ is the mean free path in the weak link and $L$ is the junction length. In the diffusive case $E_{Th} = \hbar D / L^2$ where $D$ is the diffusion constant and $L$ is the junction length. $E_{Th}$ and $l$ can in principle be extracted from the dependence of graphene normal resistance on $V_G$ \cite{Li2016,Ke2016}. However, our device has an unusual geometry and non-colinear current and voltage probes, introducing uncertainties in the determination of $E_{Th}$. Taking $l$ in the tens of nm, we estimate $E_{Th}$ to be a few hundred $\mu eV$. The gap $\Delta$ of 10 nm thick NbSe$_2$ is close to the bulk value of 1.3 meV \cite{Khestanova2018}, placing Junction A in an intermediate regime, leaning towards the long and diffusive.

It is predicted that in infinitely long metallic diffusive SNS junctions, with perfect contacts, at zero temperature $eI_C R_N=\alpha E_{Th}$ ($R_N$ is the junction normal resistance \cite{Dubos2001a}). Values near the theoretically predicted value of $\alpha = 10.82$ were seen in metal SNS junctions~\cite{Dubos2001a}, whereas in graphene $\alpha$ varies widely, reaching values as much as 100 times smaller than theory~\cite{Komatsu2012,Ke2016,Li2016,Jeong2011}. Low values of $\alpha$ are attributed to an effective Thouless energy $E_{Th}^*$, smaller that $E_{Th}$ determined by transport. This is possibly due to finite contact resistance and Andreev reflections across the N-S barrier which increase the time of junction traversal~\cite{Hammer2007,Li2016}.  
In Junction A $E_{Th}$ is of the order of $I_C R_N$, thus the proportionality factor $\alpha$ is of order unity. This is indicates an effective  $E_{Th}^*\approx0.1E_{Th}$, smaller than metallic SNS junctions and larger than previously reported diffusive graphene junctions \cite{Komatsu2012,Ke2016,Li2016,Jeong2011}.

In the long junction limit at low temperatures theory predicts~\cite{Dubos2001a}: 
\begin{equation}
e I_C R_N = \alpha_1 E_{Th} \left[ 1- b \exp\left( \frac{-\alpha_2 E_{Th}}{3.2 k_B T} \right) \right] \label{eq:temp}
\end{equation}
where $\alpha_1 = \alpha_2 = 10.82$ and $b=1.3$. Previous attempts to fit the temperature dependence in SGS junctions led to findings of $\alpha_{1,2} = 1.1-2.9$ in \cite{Jeong2011,Lee2019,Li2016}.
In Figure~\ref{fig:JJ} panel (c) we show that the temperature dependence of the critical current in Junction A fits well to an equation of this form, at low temperatures up to $T\simeq 3$K. Since we do not know the precise value of the transport $E_{Th}$, the fitting parameters are of limited quantitative value; nevertheless, assuming $E_{Th}\approx 300 \mu\textrm{eV}$, We find $\alpha_1=1.2$, $\alpha_2=2.4$ and $b=1.2$. ($\alpha_1<\alpha_2$ was also found for similar NbSe$_2$ - graphene JJ's \cite{Li_2019}). These values of $\alpha_{1,2}<10.82$ again indicate an effective $E_{Th}^*<E_{Th}$. As we will show below, measurements at parallel magnetic fields may provide another gauge for $E_{Th}^*$.
Next we observe the response of the system to the application of magnetic field $H_\perp$ perpendicular to the junction plane. Figure \ref{fig:JJ} panels (d,e,f) show $dV/dI$ as a function of $H_\perp$ and $I$, taken at three different gate voltages. The observed Fraunhofer-like pattern confirms a smooth current distribution across the junction. The apparent period is 0.4 mT. 
We compare this to the expected period $\Phi_0 / [(L+2\lambda_L)W]$, where $\Phi_0$ is the flux quantum, $L$ being the average junction length, and $W$ the junction width. 
$\lambda_L$ is the London penetration length, taken as $\lambda_L=200$ nm (known values in the literature range between $\approx 120$ nm for bulk NbSe$_2$ \cite{Callaghan2005} and 250 nm for bilayer NbSe$_2$ \cite{Talantsev2017}).
Using the above, we find the period to be $\approx 0.7$ mT - larger than the observed period, likely due to flux focusing~\cite{Suominen2017,Komatsu2012}

The junction appears to retain a homogeneous current distribution even when the Fermi energy is tuned to the Dirac point - unlike ballistic graphene devices, where transport becomes dominated by edge modes~\cite{Allen2016,Zhu2017}.  
Close scrutiny of panels (d,e,f), however, reveals discrepancies from the perfect interference pattern: lobes are not identical, and there is an asymmetry around $H_\perp = 0$. 
We suggest that this asymmetry in the interference pattern is due to spatial asymmetry in junction shape and disorder potential \cite{Rasmussen_2016,Assouline_2019,Suominen2017}. 
Additional asymmetry could arise due to the penetration of vortices into the junction area, breaking time reversal symmetry locally \cite{Golod2010,Krasnov2020}.  This will be more likely to contribute at finite $H_\parallel$.
Having confirmed that 2DJJs have transport characteristics typical to diffusive SGS junctions \cite{Heersche2007}, we turn our focus to the effect of in-plane magnetic field $H_\parallel$ on the junction.

\begin{figure}
    \centering
    \includegraphics[width=0.9\textwidth]{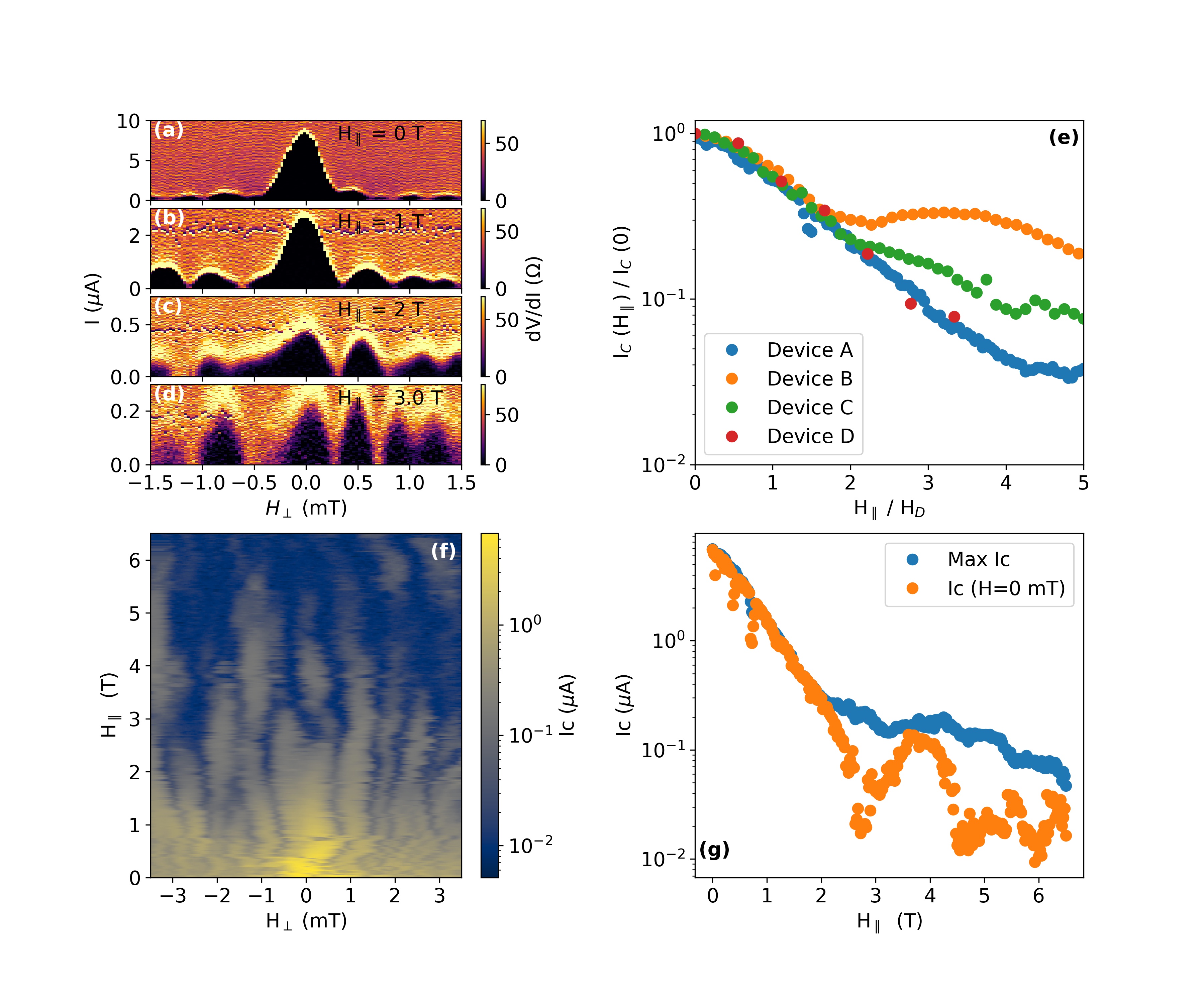}
    \caption{ \textbf{a., b., c., d.} Differential resistance of Junction A as a function of bias current and external perpendicular magnetic field, taken with applied in-plane magnetic field of 0, 1 T, 2 T, and 3 T, respectively. All measurements conducted with $V_G$ = 20 V. \textbf{e.} Parallel field dependence of the maximal critical current for Junctions A, B, C, D. $I_C$ is normalized to $I_C (H_\parallel = 0, H_\perp =0)$ and the in-plane field is normalized by a Junction-specific decay field $H_D$. Each value is extracted from a 2D scan of $R(V,H_\perp)$ at a given $H_\parallel$ and is defined as the maximal $I_C$ obtained in each scan. $H_D$ = 0.6, 0.16, 0.4, and 0.9 T for Junctions A,B,C,D respectively. The field at which exponential decay slows, $H_T$, is indicated by a dotted line for junctions A,B,C \textbf{f.}  $I_C$ of Junction A as a function of $H_\parallel$ and $H_\perp$. The curves were shifted to correct for sample misalignment, and were then aligned to be as continuous as possible. Logarithmic color-scale.  \textbf{g.} Dependence of the maximal $I_C$ (blue) and of $I_C$ at $H_\perp=0$ (orange), extracted from panel (f).} 
    \label{fig:field}
\end{figure}

\begin{figure}[h]
  \centering
  \includegraphics[width=1.0\linewidth]{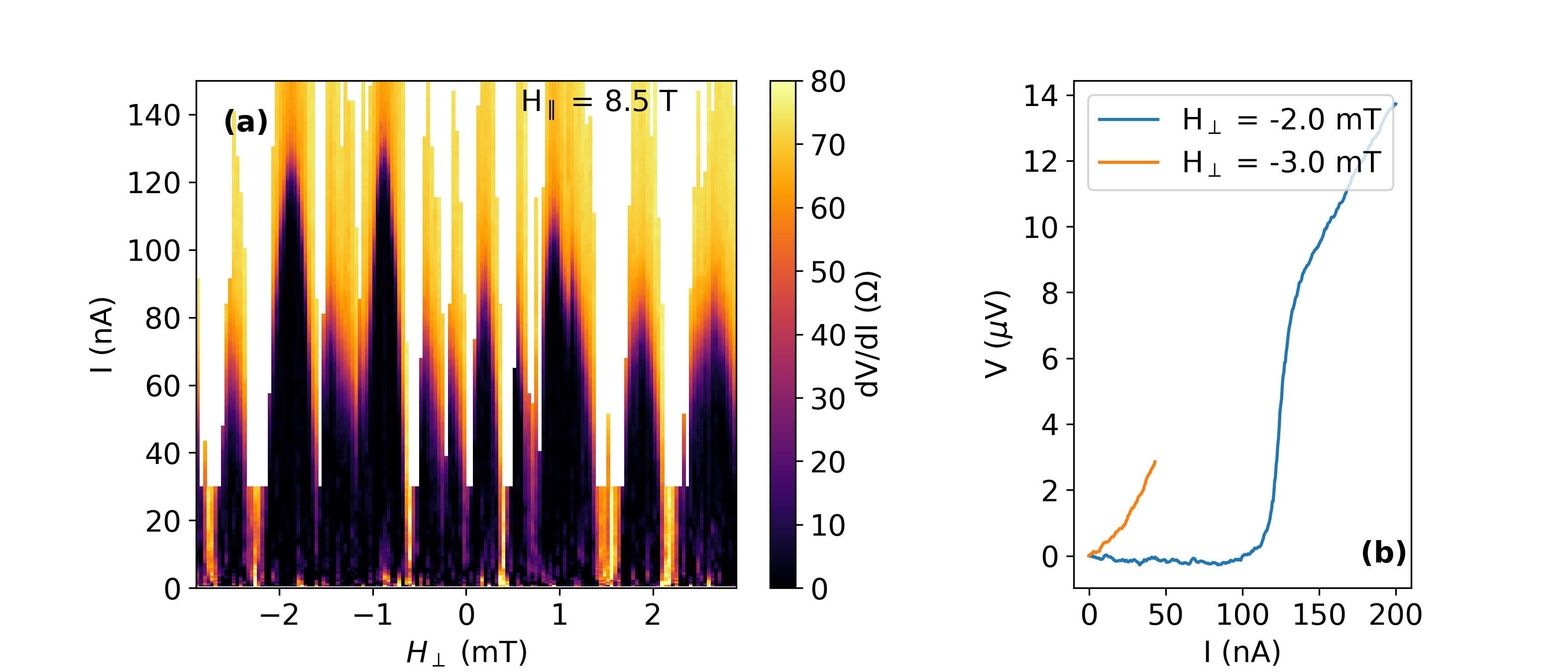}
  \caption{\textbf{High field supercurrent} \textbf{(a)} Interference pattern of Junction A at parallel field of 8.5 T shows clear lobes of zero resistance (data is from a different cool down than \ref{fig:field}). \textbf{b} I-V curve from (a) where the critical current is maximal (blue) and minimal (orange). The measurement was set to stop when normal state transport is observed. }
  \label{fig:s_highfield}
\end{figure}

Since the junction is sensitive to out-of-plane fields $H_\perp$ on the scale of a few hundreds of $\mu$T, extreme care is needed when aligning $H_\perp$ and $H_\parallel$ in our vector magnet to the sample geometric tilt. We do this by measuring the out-of-plane interference pattern, at any given $H_\parallel$.
At low fields, of up to 1.5 T in Junction A, the interference pattern shows a clearly distinguishable central lobe (Fig.~\ref{fig:field}a,b), allowing for unambiguous identification of the absolute field orientation. 
At higher $H_\parallel$ this is no longer possible: the central lobe is suppressed to the same magnitude of the side lobes (Fig.~\ref{fig:field}c,d). 
This SQUID-like supercurrent distribution may be retained up to high parallel field. Junction A for example retains its critical current at a field of $H_\parallel=8.5T$, showing a SQUID like lobe structure as a function of H$_\perp$ (Fig.~\ref{fig:s_highfield} (a)). The voltage as a function of current curve shows a clear transition from superconducting to normal state at a critical current of I=100 nA for H$_\perp=$-2 mT (panel (b)).

Once the central lobe is no longer distinguishable, there is in general no straightforward indication to the true position of $H_\perp = 0$.
Lacking this identification, we take the maximal $I_C(H_\perp)$ (hence $I_C^{max}$) as a measure for the junction critical current at each $H_\parallel$. 
We find that $I_C^{max}(H_\parallel)$ exhibits an exponential-like decay - corresponding to the suppression of the central lobe seen in Fig~\ref{fig:field}a-c. Normalizing to $I_C^{max}(H_\parallel = 0)$, we plot $I_C^{max}(H_\parallel)$ in panel (e) for Junctions A,B,C,D (all have monolayer graphene weak links, except Junction B which is bilayer graphene). To see the universality of the decay of $I_C(H_\parallel)$, we normalize it by a junction-specific decay field, $H_D$. 
The universal decay in $I_C^{max}$ persists up to a second characteristic field scale, $H_T$, where $I_C^{max}$ stabilizes to the critical current of the side lobes. Depending on the sample, at $H > H_T$ the exponential decay in $I_C^{max}$ either becomes moderate, or even turns into a small increase. For Junctions A,B,C shown in panel (e) this field is given by $H_T =$ 2.4 T (4 $H_D$), 0.24 T (1.5 $H_D$), and 0.8 T (2 $H_D$) respectively. In Junction D there are not enough data points to quantify this field. 
The junctions thus evolve to a SQUID-like lobe structure at finite, yet device-dependent $H_\parallel$ (see Supplementary Section 3 \cite{supp}). The Zeeman effect in a uniform junction predicts universal decay with $H_\parallel$. Deviation from universal behavior at $H_T$ could be a result of ripples or junction non-uniformity as we will discuss.

We now turn our attention to Fig.~\ref{fig:field}(f), which depicts the evolution of $I_C$ vs. $H_\perp$ and $H_\parallel$ in Junction A. In this junction we were able to track the evolution of the interference pattern up to $H_\parallel = 6.5$ T, aligning the $I_C(H_\perp)$ curves as explained in Supplementary Section 1 \cite{supp}, and thus obtaining the map shown in panel (f)\footnote{data in \ref{fig:field} panels (a)-(d) comes from a different measurement than panel (f), taken on the same device, and show a slightly different lobe structure}. The magnitude of $I_C(H_\parallel,H_\perp = 0)$, plotted in panel (g), shows a suppression and recovery pattern. This data is reminiscent of suppression-recovery patterns seen in superconductor-ferromagnet-superconductor (SFS) junctions~\cite{Kontos_2002,Guichard_2003,Blum_2002,Oboznov_2006,Shelukhin_2006,Robinson_2006} and in 2D systems~\cite{Hart_2016,Chen_2018,Li_2019}, where it is interpreted as a $0-\pi$ transition.

The salient features of the data are therefore (a) exponential decay of the critical current at low field (b) saturation of the critical current at intermediate fields (c) lobe structure transition from Fraunhofer-like to SQUID-like, and (d) vanishing and reappearing of the central lobe critical current in Device A. 
In what follows, we discuss the physics in our 2DJJ by considering both the parallel field-tunable Zeeman splitting of the graphene band structure, and the orbital effect of out-of-plane ripples in the graphene \cite{Fyhn2020}. 

Lacking an intrinsic spin-orbit coupling, graphene dispersion is affected by magnetic field only through Zeeman splitting, where the Zeeman energy is analogous to the exchange interaction in SFS JJs \cite{Buzdin2005, Fyhn2020}. In the latter, the superconducting order parameter in the ferromagnetic layer varies as the product of an an exponential decay and an oscillatory term:
\begin{equation}
    \psi(x)= \psi_i\exp(-k_1 x) \cos(k_2 x)
\end{equation}
where $\psi(x)$ is the order parameter at the position $x$ along the junction, $\psi_{i}$ is the order parameter at the superconducting lead, and $k_1$, $k_2$ are the inverse characteristic length scales associated with the decay and oscillation. In the diffusive limit, they are both given by $1/k_1, 1/k_2 = \sqrt{L^2 E_{Th}/{E_Z}}= \sqrt{2D / g \mu_B H_\parallel}$ where D is the diffusion coefficient.

The order parameter thus experiences a decay accompanied by oscillation, with zeros occurring periodically when $Lk_2=\pi/2+n\pi$, or $E_Z=(\pi/2 +n\pi)^2 E_{Th}$. This behavior of the order parameter leads to an oscillatory decay of the critical current of the junction.
Following this intuition, critical current of an SGS junction in a parallel magnetic field is thus expected to undergo an exponential suppression at low fields - in agreement with our observations. The oscillatory component of the wave function leads to a  \mbox{0-$\pi$}~transition: a change of the equilibrium phase difference between the two superconducting leads, accompanied by a reversal of the supercurrent.

\begin{figure}
    \centering
    \includegraphics[width=0.7\textwidth]{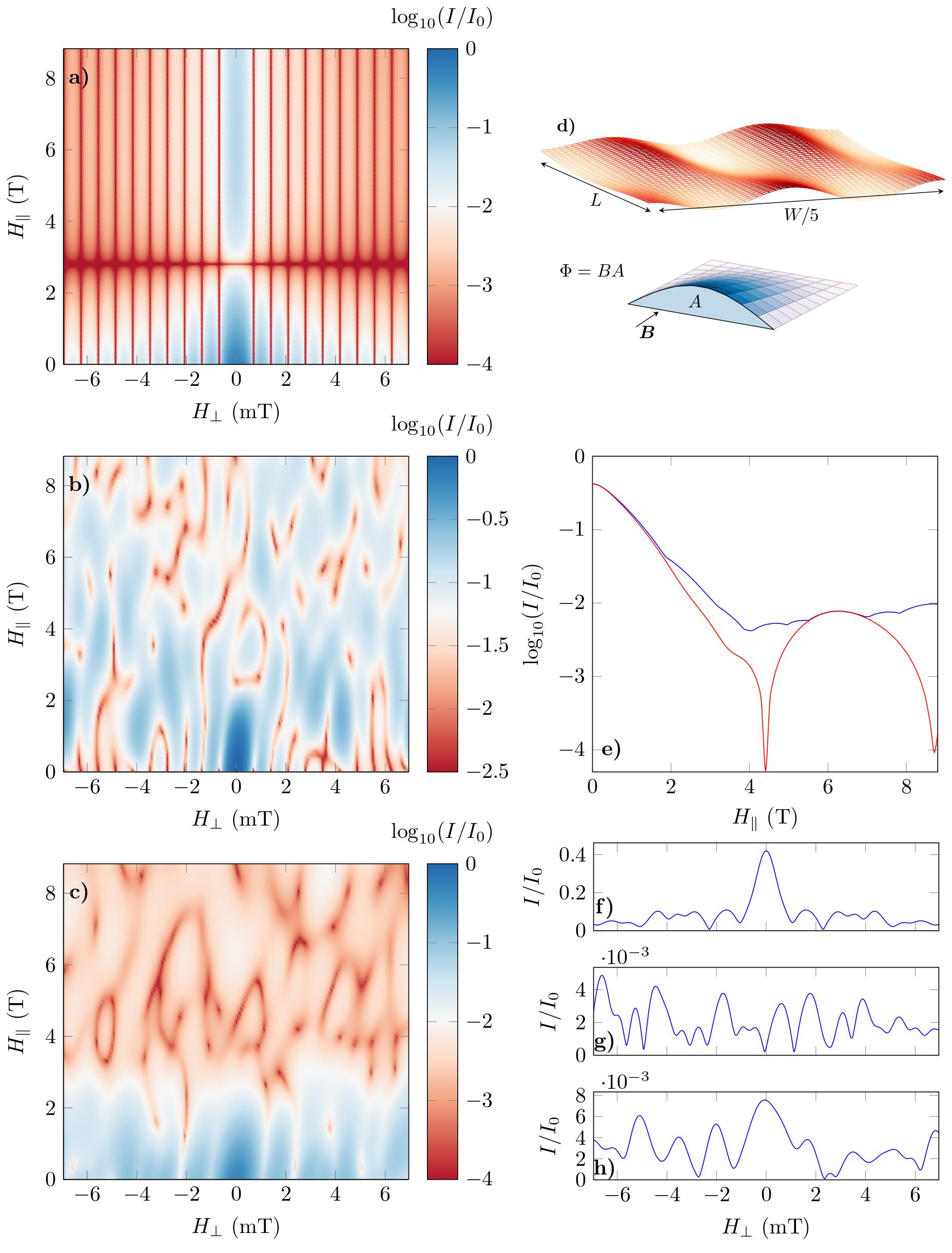}
    \caption{   ~\textbf{a.-c.} Calculated critical current $I_C$ with logarithmic color-scale as a function of $H_\perp$ and  $H_\parallel$ \textbf{a.} simulated Zeeman effect with $E_{Th}= 64$ $\mu eV$, and a rectangular junction of dimensions $L=214$ nm, $W=4.69$ $\mu$m without ripples \textbf{b.} a rectangular junction with ripples, disregarding the Zeeman effect and \textbf{c.} our measured junction contour with varying L, $E_{Th} = 64\mu eV$, Zeeman effect and ripples \textbf{d.} Top: ripple profile used to generate the maps in (b),(c). Note that the actual aspect ratio is around W/L=20 Bottom: illustration of a long wavelength ripple which could give rise to a zero in I$_c$ at low fields. \textbf{e.} $I_C$ at $H_\perp=0$ (red   ) and maximal $I_C$ for all $H_\perp$ (blue) vs. $H_\parallel$, line cuts taken from the simulation in (c). \textbf{f.-h.} $I_C$ vs. $H_\perp$ for $H_\parallel=0,4.1$ and $6.2$ T, line cuts from the simulation in (c)}
    \label{fig:theory}
\end{figure}

Using the analytical solution of the Usadel equations in an SGS junction \cite{Fyhn2020} to qualitatively model our system, we calculate the critical current as a function of $H_\parallel$ and $H_\perp$ specifically for Junction A.
We assume the junction length $L = 214$ $nm$ and width $W = 4.69 \mu$m (average dimensions taken from the SEM measurement shown in Fig. \ref{fig:fab} (c)). Results are shown in Fig.  \ref{fig:theory} panel (a). The assumed uniform supercurrent reversal manifests in the suppression of all lobes, corresponding to the disappearance of the uniform supercurrent throughout the junction at a numerically determined transition field ~\cite{Fyhn2020}:
\begin{equation}
    H_\parallel \approx \frac{2.5 E_{Th}}{0.5g \mu_B}=\frac{5D\hbar}{g \mu_B L^2} 
\end{equation}
From the experimentally observed transition field of 2.8 T,  assuming $g=2$ we find $E_{Th}=64 \mu eV$. This falls between the order of magnitude expected for $E_{Th}$ of hundreds of $\mu eV$ which we extract from normal regime transport properties, and $E_{Th}^*$ of tens of $\mu eV$ extracted from $I_CR_N$. Recalling that a lower effective $E_{Th}$ has been attributed to Andreev reflection across an imperfect S-N interface, we point out that there is to the best of our knowledge no theory addressing how this would effect the Zeeman physics in the junction. 

In the data in Fig. \ref{fig:field} (f) we find that high order lobes are retained while the zero lobe, representing the average supercurrent, is suppressed. This indicates that the supercurrent is nonuniform. 
When multiple transport channels are present, they may carry positive and negative supercurrents which cancel out at H$_\parallel$ = 2.8 T where the central lobe vanishes. In this regime the other lobes of the interference pattern, measuring higher moments of the supercurrent with respect to the out-of-plane field, should not in general disappear. This phenomenon was seen in SFS JJs with a non-uniform ferromagnetic barrier, leading to a similar interference pattern \cite{Kemmler2010,Frolov2006,Pfeiffer2008,Weides2006}.  
Non-uniformity in supercurrent reversal can arise from local variation in $E_{Th}$, since regions with lower $E_{Th}$ will undergo stronger suppression due to $E_Z$. Such variation in $E_{Th}$ can arise from varying junction length, as well as from local variations in contact transparency. Additionally, it could be a consequence of charge disorder, locally affecting the diffusion constant.
However, the observed SQUID-like interference pattern can only be reproduced by an $E_{Th}$ profile which sharply favors edge transport.

We now turn to the orbital effects associated with the locally varying perpendicular components of H$_\parallel$. These variations may be caused either by graphene height variations, or by disruptions to the parallel field due to the Meissner effect, which diverts flux lines around the superconducting electrodes (flux-focusing). Because both ripples and flux-focusing give rise to a spatially varying
perpendicular field component, their effect on the electric current is similar. For concreteness we give an in-depth discussion of the ripple scenario, but note that underlying mechanism could in principle also be flux-focusing.

Using the same model discussed previously, we distinguish between the effects of short and long wavelength ripples \cite{Fyhn2020}.  
Short ripples as seen in microscopy studies of graphene on SiO$_2$ are typically $\approx 0.3$ nm peak-to-peak with a correlation length of 10-30 nm~\cite{Ishigami2007,Geringer2009,Cullen2010,Xue2011}. Long ripples have a wavelength larger than the junction dimensions.
Intuitively, one may gauge the effect of a ripple by calculating the flux accumulated within an area defined by the ripple lateral cross section, illustrated in Fig.~\ref{fig:theory} (d). To induce a full current suppression and revival at $H_\perp=0$, a ripple within the junction has to accumulate a single flux quantum due to the parallel field, according to the equation:
\begin{equation}
    H_\parallel=\frac{\Phi_0}{\eta \lambda}
\end{equation}
where $\eta$ is the average ripple amplitude within the junction, and $\lambda$ is the wavelength (or the limiting junction dimension if the ripple extends beyond the junction). For the typical short wavelength ripple seen in graphene on SiO$_2$,  parallel fields of order 50 T are required to obtain an entire flux quantum within a ripple. However the cumulative effect of many such ripples causes a faster decay of the critical current which can create exponential-like behavior, similar to the Zeeman effect \cite{Fyhn2020}.

2DJJs in a parallel field are highly sensitive to long-wavelength height variations ~\cite{Fyhn2020}. 
In our experimental geometry, with junction width $W \approx 4.7 \mu m$, it is possible to consider a ripple of length $\lambda \simeq W$.
As a long wavelength feature accumulates much more flux, it is possible to reach a flux quantum given a few T parallel field and a small height variation of $\eta \approx 0.1$ nm within the junction.
We note that based on AFM and STM studies, it is difficult to tell whether such sub nm height variations are present over micron length scales. Such geometry is physically conceivable due to strain or curvature of the substrate, and cannot be ruled out. 
We show the simulated supercurrent in a sample ripple configuration containing ripples in Fig.~\ref{fig:theory} (b). The simulation reproduces the features of the data highlighted previously - exponential decay followed by saturation, lobe structure transition, and a critical current dip at around B = 3 T. The specifics of these features, such as the location and sharpness of the critical current dip, vary with different ripple configurations, however many different patterns can produce qualitatively similar results (see Supplementary Section 2 \cite{supp}). Fig.~\ref{fig:theory} (d) illustrates the specific ripple profile used to obtain the map in (b). The simulation does not include ripples of wavelength smaller than around 100 nm. These in general cause a sharper decay of critical current with parallel field \cite{Fyhn2020}. 

Since we expect Zeeman and ripple effects to coexist, we present a compound simulation which considers them both (Fig. \ref{fig:theory} (c)). This simulation also accounts for varying $E_{Th}$ due to variation in junction length as extracted from the SEM data presented in Fig. \ref{fig:fab} (c). In the case of varying junction length our analytical model is not rigorous, but it does give a qualitative approximation. 
As we see in panel (e), the simulation reproduces the exponential decay, suppression and recovery of $I_C(H_\parallel,H_\perp=0)$. The lobe structure at $H_\parallel=0,4.1,6.2$ T (panels (f)-(h)), exhibits the experimentally observed transition between Fraunhofer-like and SQUID-like profiles. 

\begin{figure}[h]
    \centering
    \includegraphics[width=0.9\textwidth]{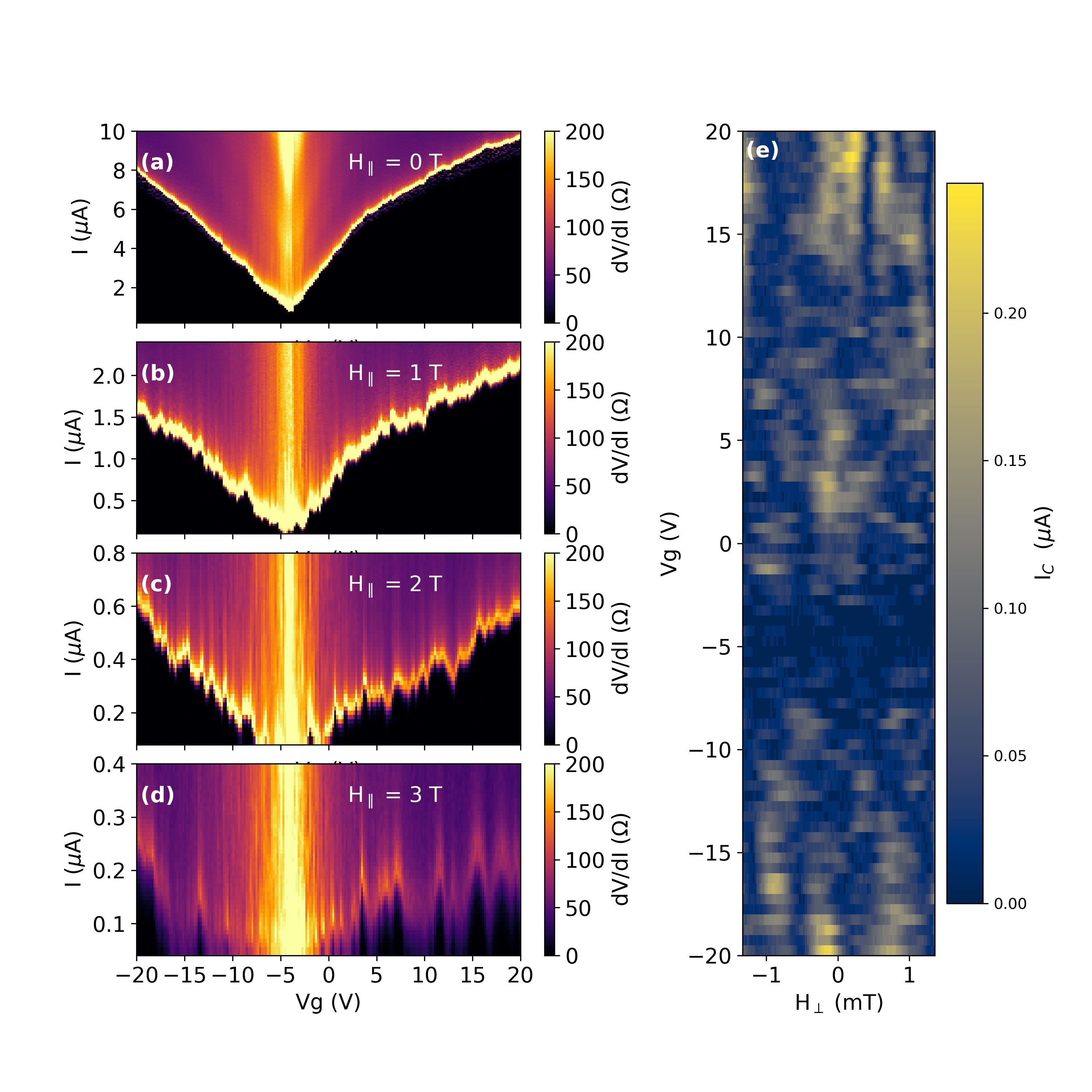}
    \caption{ \textbf{ a.-d.} Differential resistance of Junction A as a function of bias current and gate voltage, taken with applied in-plane magnetic field of 0, 1 T, 2 T, and 3 T respectively. All measurements conducted with $H_\perp = 0$.    \textbf{e.} $I_C$ vs. $H_\perp$ and $V_G$ for Junction A, taken with parallel field $H_\parallel$ = 3 T.  } 
    \label{fig:gate}
\end{figure}

Finally, we observe how the application of $H_\parallel$ affects the gate dependence of the critical current (Fig. \ref{fig:gate}). At zero field $I_C$ varies smoothly with Vg (panel a), leading to a nearly constant $I_C R_N$ product away from the Dirac point. Upon increasing $H_\parallel$, $I_C$  fluctuates with $V_G$ (panels b-d) leading at $H_\parallel = 3$ T to patterns of decay and revival of $I_C(V_G)$. 
Observing the evolution of the interference pattern with $V_G$ at the same field reveals a qualitative change in the number of visible lobes and in their positions (panel (e)). 

The observed gate dependence of the interference pattern shows that at $H_\parallel$ around the suppression-recovery field of 2.8T, the junction enters a new regime where the critical current survives in patches at fluctuating gate values. Similar phenomenology has been observed in ballistic graphene JJs at high perpendicular field, and attributed to chaotic billiards due to cyclotron orbits reflecting from the graphene edge \cite{Shalom2016}. However the physics in our regime is different, since the junction is diffusive and $B_\perp \approx 0$. Within the Zeeman effect interpretation, it could be due to local gate driven fluctuations around the 0-$\pi$ transition as in \cite{Ke_2019}. Alternatively, when ripples become important, changing gate could change the resulting interference pattern. There could also be a gate-dependent effect in the contact region between the graphene and NbSe$_2$. 
In any case, clearly the current flow distribution in this regime depends strongly on graphene Fermi energy. This could be linked to local charge conditions such as the disorder potential landscape, however, the lobe structure continues to evolve when the graphene is at high carrier densities, where disorder potential should be screened.

We conclude that the 2DJJ architecture allows the study of graphene Josephson junctions at high parallel magnetic fields, where supercurrent is sensitive to both the Zeeman effect and sub-nm graphene height variations. Junction currents evolve from a Fraunhofer-like to a SQUID-like interference pattern.
We observe a supercurrent suppression and recovery feature which may be associated with a Zeeman driven $0-\pi$ transition, or with the accumulation of a single flux quantum within a $\mu m$-wavelength ripple.
While in the present measurements it is difficult to distinguish between the two effects, future experiments, with graphene placed on hBN, are expected to suppress the ripple contribution. In the future it will be interesting to consider devices of the 2DJJ architecture utilizing different 2D materials as contacts and weak links. For example, devices where graphene inherits a spin-orbit term from a TMD substrate. The combination of significant spin-orbit and high parallel magnetic fields in the context of a Josephson junction, could give rise to topological effects~\cite{Wakamura2020}.

The authors wish to thank M. Aprili, Y. Oreg, A. Stern, F. Pientka, and A. Di Bernardo for illuminating discussions. This work was funded by a European Research Council Starting Grant (No. 637298, TUNNEL), Israeli Science Foundation grant 861/19, and BSF grant 2016320. T.D. and A.Z. are grateful to the Azrieli Foundation for Azrieli Fellowships. K.W. and T.T. acknowledge support from the Elemental Strategy Initiative conducted by the MEXT, Japan ,Grant Number JPMXP0112101001,  JSPS KAKENHI Grant Number JP20H00354 and the CREST(JPMJCR15F3), JST.

\section{Supplementary Section: Measurement and field alignment procedure}
The system studied here, an SNS junction using graphene as the weak link between  NbSe$_2$ SC leads, is very sensitive to the presence of perpendicular fields on the scale of hundreds of $\mu$T. When applying parallel magnetic field, perpendicular field can also be present, either from a small misalignment of the sample within the magnet, or from the presence of vortices and trapped magnetic flux in the leads, in the junction or in the magnet itself. At low parallel magnetic fields, the interference pattern of the supercurrent with the application of perpendicular field shows a clear maximum at zero applied field. Thus, it is possible to track the shift of this maximum with applied parallel field and find the sample misalignment. In the measurements reported in this work we have done so, and found the required amount of perpendicular field to compensate for this effect. The interference patterns reported here are always with respect to the corrected zero perpendicular field.

At higher magnetic fields, this correction is not enough, as remnant field, coming from vortices in the leads and trapped flux in the magnet affects the sample. To correct for that, we assume that a small change in the parallel field should not create a large change in the interference pattern of the junction. Based on this assumption, when analyzing the data, we use the following alignment procedure: We shift the interference pattern measured at a given parallel field by a some amount of perpendicular field. We calculate the sum of the squared differences between an this pattern and the pattern measured in the previous parallel field. We repeat this for a series of shifts and find for which shift this difference is minimal. We choose this shift as the correct alignment for the pattern, and repeat with the pattern taken at the next step of the parallel field.

The above procedure was utilized for junction A. For other junctions, we could not determine the orientation of $H_\perp$=0T at high values of H$_\parallel$ due to jumps in the interference pattern. Therefore we used the maximal critical current  $I_C^{max}$ as an indicator of the junction critical current.

\begin{figure}[h]
  \centering
  \includegraphics[width=1.0\linewidth]{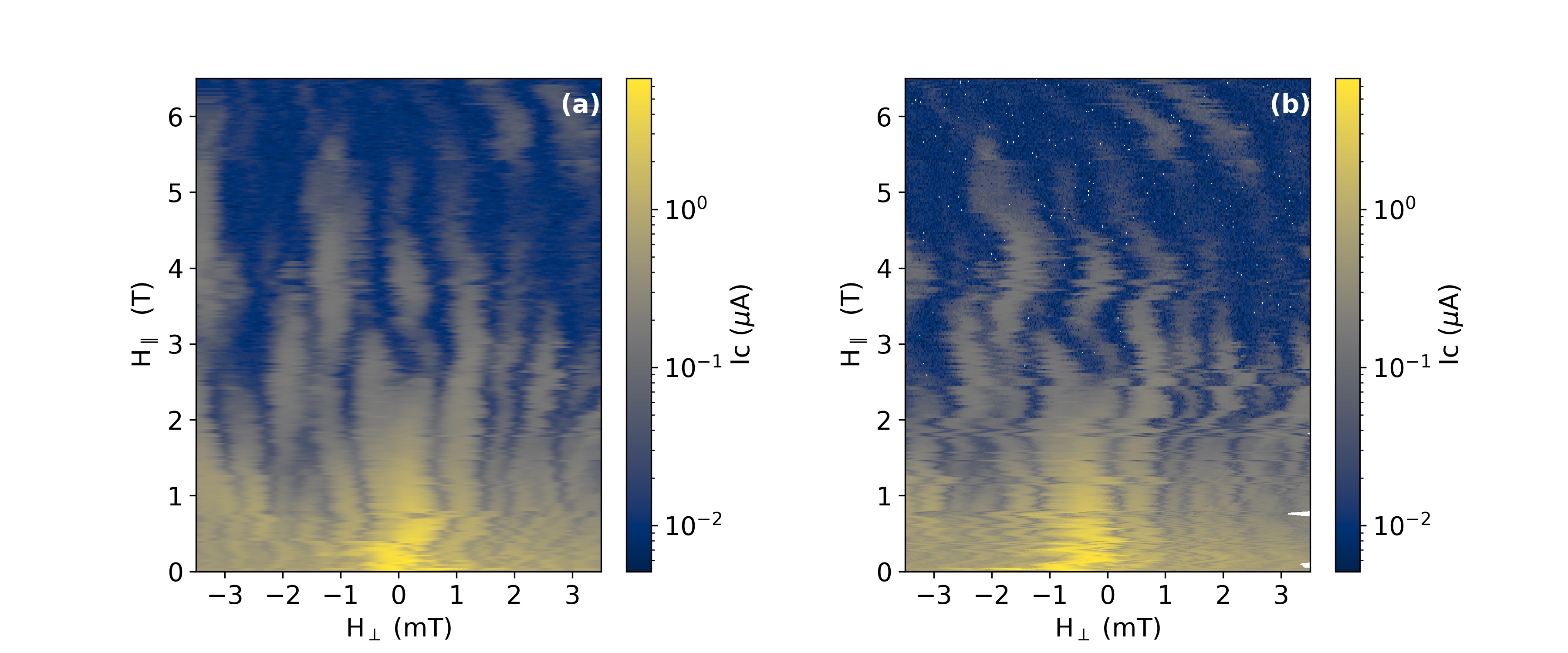}
  \caption{\textbf{(a)} The 2D map of I$_C$ of Junction A as a function of $H_\parallel$ and $H_\perp$, aligned using the full alignment procedure of minimizing the squared difference between line-scans and shown also in main text Fig. 2(f). \textbf{(b)}  The same data shown when only geometrical misalignment was taken into consideration.}
  \label{fig:s_alignment}
\end{figure}

\pagebreak

\section{Supplementary Section: Simulation of ripples in graphene}

We assume the presence of height variations in graphene, and in the absence of the Zeeman effect (taking $E_{Th}>>E_Z$), theoretically calculate the critical current expected for Junction A, shown in main text Fig. 3 panel (b). We calculated this for eight different ripple profiles, generated by taking a sum of a random number of sine functions, all with randomized wavelength, amplitude and angular offsets. The wavelengths were taken from an exponential distribution (meaning that longer wavelengths are more probable) while the rest of the parameters were taken from uniform distributions. Qualitative features such as a transition in the lobe structure from Fraunhofer-like to SQUID-like, decay of the critical current and a suppression and reappearance of I$_C$(H$_\perp$=0) appear for several of the randomly generated profiles. The specific result we chose to present in the main text was generated from the following ripple profile:

\begin{equation}
    \eta (x,y) = 1.25 \sin{64.6y+1.7164}\sin{3.5x-0.7}+0.45\sin{37.3y+2.136}\sin{2x}
    +1.3\sin{2y+0.3224}\sin{0.1x+0.03}
\end{equation}

\begin{figure}[h]
  \centering
  \includegraphics[width=1.0\linewidth]{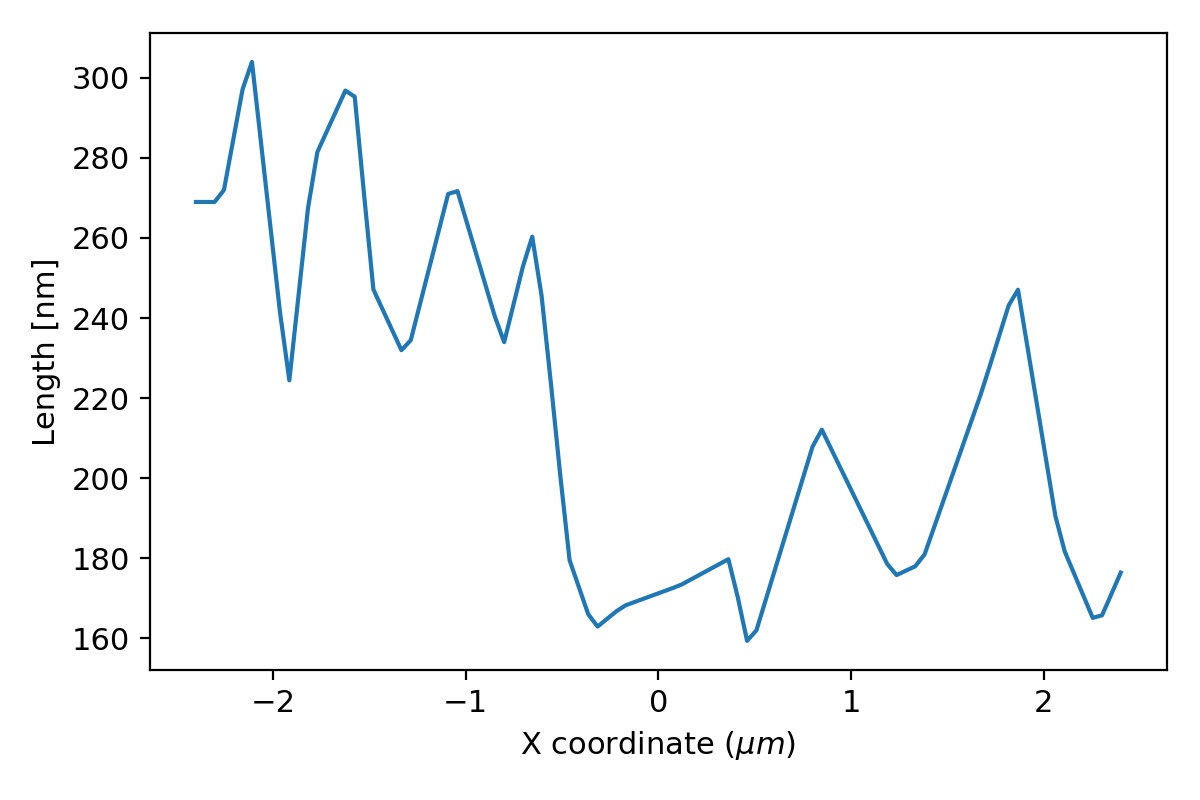}
  \caption{The junction length profile of Junction A as measured by SEM.} 
  \label{fig:ripples}
\end{figure}

Ripple amplitude is given in nm, while the coordinates x,y are normalized and range between -0.5,0.5. The profiles were chosen with wavelengths much larger than 50 nm, in order to accommodate the use of the wide-junction approximation~\cite{bergeret2008}. The junction length profile used for the combined simulation of Zeeman, ripple and junction geometry effects presented in main text Fig. 3 panel (c), is illustrated in Fig.~\ref{fig:ripples}.

\section{Supplementary Section: Additional junctions}

We present critical current measurements from devices B,C,D, showing $I_C$ vs $H_\perp$ (Junction B has a weak link of bilayer graphene, the rest are monolayer). This exhibits a Fraunhofer-like interference at low $H_\parallel$ and a more SQUID-like lobe structure, with lobes having a similar maximal I$_C$, at higher $H_\parallel$ for each device. Junction D has a  nonuniform geometry with multiple graphene weak links, making it difficult to directly interpret the interference pattern. Roughly, the lobe structure at $H_\parallel$=0 has a large area non-uniform component leading to high frequency lobes, modulated by a slow decay due to the uniform current in an individual small junction. The later modulation becomes more uniform as $H_\parallel$ increases, as in the other devices. 

\begin{figure}[h]
  \centering
  \includegraphics[width=1.0\linewidth]{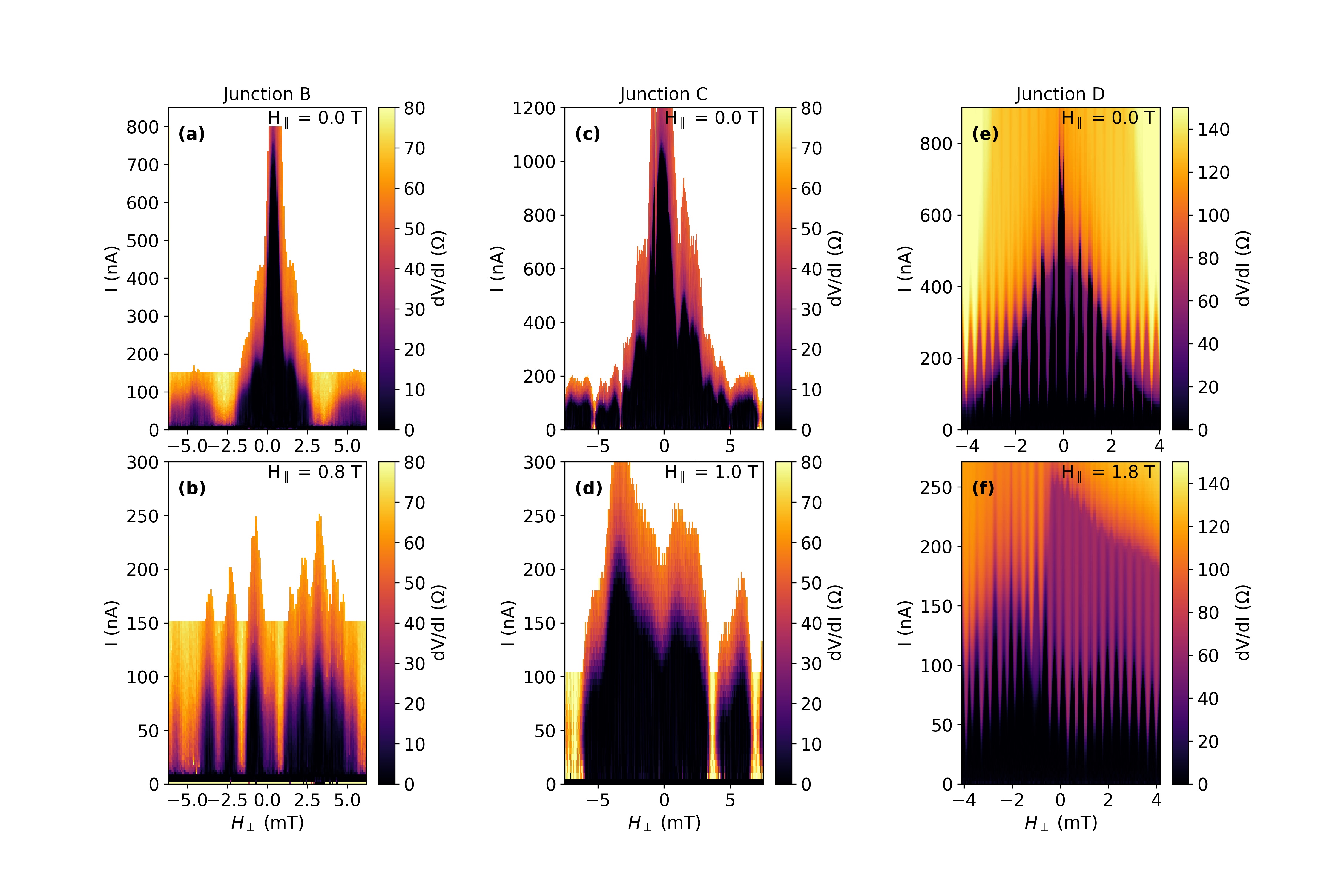}
  \caption{\textbf{Interference patterns from Junctions B,C,D} \textbf{(a-c)} Interference patterns of Junctions B,C,D respectively at $H_\parallel$=0 T showing a pronounced central lobe in each case \textbf{(d-e)} Interference patterns at various values of non-zero $H_\parallel$ (indicated in the panel) for device B,D,C respectively. These patterns have all lobes with similar maximal values of $I_C$, indicating a non-uniform current distribution  }
  \label{fig:s_BCD}
\end{figure}

\section{Supplementary Section: Fabrication Methods}

To fabricate a graphene - NbSe$_2$ Josephson junction (JJ), we first exfoliate graphene on markered SiO$_2$ and locate suitable flakes. Next, NbSe$_2$ is exfoliated on PDMS gel and examined to find flakes which are a few layers thick and have an observable crack, less than 500 nm wide. Chosen NbSe$_2$ flakes are transferred onto graphene using the viscoelastic dry-transfer method \cite{Gomez2014}. A few nm thick hBN flake may then be transferred over the crack to serve as a protective layer and a potential top-gate dielectric. The NbSe$_2$ flake is contacted with standard e-beam lithography using Ti/Au contacts. Prior to evaporation of contacts surface oxide was removed using Argon ion milling. Four-probe  measurements were conducted in a dilution cryostat with a base temperature of 30 mK (see main text Fig. 1).

\bibliographystyle{ieeetr}

\end{document}